\documentclass[12pt,draft,notref,notcite]{article}

\usepackage{color,longtable}
\usepackage{amssymb,graphicx}

\newcommand{\be}{\begin{eqnarray}}
\newcommand{\ee}{\end{eqnarray}}

\usepackage{amsfonts,amsmath}
\usepackage{latexsym}

\def\cH{{\cal H}}

\def\a{{\alpha}}
\def\b{{\beta}}

\def\E{E_{10}}
\def\K{K(E_{10})}
\def\KE{E_{10}/K(E_{10})}

\begin{document}

\begin{center}

{\bf \Large Cosmological Singularities and a Conjectured Gravity/Coset Correspondence}\footnote{Contribution to ``String Theory and Fundamental Interactions'' -- in celebration of Gabriele Veneziano's 65th birthday -- eds. M. Gasperini and J. Maharana, Springer-Verlag, Heidelberg, 2007.}

\bigskip

Thibault Damour

\medskip

{\sl Institut des Hautes Etudes Scientifiques, 35 route de Chartres,  \\ F-91440 Bures-sur-Yvette, France}
\end{center}

\vspace{1cm}

\begin{minipage}{12cm}
\textbf{Abstract:}
We review the recently discovered connection between the Belinsky-Khalatnikov-Lifshitz-like ``chaotic'' structure of generic cosmological singularities in eleven-dimensional supergravity and the ``last'' hyperbolic Kac-Moody algebra ${\E}$. This intriguing connection suggests the existence of a hidden ``correspondence'' between supergravity (or even $M$-theory) and null geodesic motion on the infinite-dimensional coset space $\KE$. If true, this gravity/coset correspondence would offer a new view of the (quantum) fate of space (and matter) at cosmological singularities.
\end{minipage}

\vspace{1cm}

\section {Introduction}

It is a pleasure to participate in the celebration of the seminal accomplishments of Gabriele Veneziano. I will try to do so by reviewing a line of research which is intimately connected with several of Gabriele's important contributions, being concerned with the cardinal problem of String Cosmology: the fate of the Einstein-like space-time description at big crunch/big bang cosmological singularities. Actually, the work, described below started as a by-product of the string cosmology program initiated by M. Gasperini and G. Veneziano \cite{Gasperini:2002bn}.  While collaborating with Gabriele on the possible  birth of ``pre-big bang bubbles'' from the gravitational-collapse instability of a {\em generic} string vacuum made of a stochastic bath of incoming gravitational and dilatonic waves \cite{Buonanno:1998bi}, an issue raised itself : what is the structure of a {\em generic} spacelike (i.e. big crunch or big bang) singularity within the effective field theory approximation of (super-) string theory (when keeping all fields, and not only the metric and the dilaton). The answer turned out to be surprisingly complex, and rich of hidden structures. It was first found \cite{Damour:2000wm,Damour:2000th} that the general solution, near a spacelike singularity, of the massless bosonic sector of all superstring models  ($D = 10$, IIA, IIB, I, HE, HO), as well as that of $M$ theory ($D = 11$ supergravity), exhibits a never ending oscillatory behaviour of the Belinsky-Khalatnikov-Lifshitz (BKL) type \cite{Belinsky:1970ew}. However, it was later realized that behind this seeming entirely {\em chaotic} behaviour there was a {\em hidden symmetry structure} \cite{hep-th/0012172,hep-th/0103094,hep-th/0207267}. This led to the conjecture of the existence of a hidden equivalence (i.e. a {\em  correspondence}) between two seemingly very different dynamical systems: on the one hand, $11$-dimensional supergravity (or even, hopefully, ``$M$-theory''), and, on the other hand, a {\em one-dimensional} $\KE$ nonlinear $\sigma$ model, i.e. the geodesic motion of a massless particle on the infinite-dimensional coset space\footnote{Here $\K$ denotes the (formal) ``maximal compact subgroup'' of the hyperbolic Kac-Moody group $E_{10}$.} $E_{10} / K (E_{10})$ \cite{hep-th/0207267}. The intuitive hope behind this conjecture is that the BKL-type {\it near spacelike singularity limit} might act as a tool for revealing a hidden structure, in analogy to the much better established AdS/CFT correspondence \cite{hep-th/9905111}, where the consideration of the {\it near horizon limit} of certain black $D$-branes has revealed a hidden equivalence between 10-dimensional string theory in AdS spacetime on one side, and a lower-dimensional CFT on the other side. If the (much less firmly established) ``gravity/coset correspondence'' were confirmed, it might provide both 
the basis of a new definition of $M$-theory, and a description of the ``de-emergence'' of space near a cosmological singularity (see \cite{DNGRF07} and below).

\section{Cosmological billiards}

Let us start by summarizing the BKL-type analysis of the ``near spacelike singularity limit'', that is, of the asymptotic behaviour of the metric $g_{\mu\nu}(t, \textbf{x})$, together with the other fields (such as the 3-form $A_{\mu \nu\lambda} (t, \textbf{x})$  in supergravity), near a singular hypersurface. The basic idea is that, near a spacelike singularity, the time derivatives are expected to dominate over spatial derivatives. More precisely, BKL found that spatial derivatives introduce terms in the equations of motion for the metric which are similars to the ``walls'' of a billiard table \cite{Belinsky:1970ew}. To see this, it is convenient \cite{hep-th/0212256} to decompose the $D$-dimensional metric $g_{\mu\nu}$ into non-dynamical (lapse $N$, and shift $N^i$, here set to zero) and dynamical ($e^{- 2 \beta^a}$, $ \theta^{a}_i)$ components. They are defined so that the line element reads 
\begin{equation}
\label{eq1}
ds^2 = - N^2 dt^2 + \sum_{a = 1}^{d} 
e^{-2 \beta^{a}} \theta^a_i \theta^a_j  dx^i dx^j. 
\end{equation}
Here $d \equiv D- 1$ denotes the spatial dimension ($d = 10$ for SUGRA$_{11}$, and $d = 9$ for string theory), $e^{- 2 \beta^a}$ represent (in an Iwasawa decomposition) the ``diagonal'' components of the spatial metric $g_{ij}$, while the ``off diagonal'' components are represented by the $\theta^{a}_i$, defined to be upper triangular matrices with 1's on the diagonal (so that, in particular, $\det \theta = 1$).

The Hamiltonian constraint, at a given spatial point, reads (with $ \tilde{N} \equiv N / \sqrt{\det g_{ij}}$ denoting the ``rescaled lapse'')
\begin{eqnarray}
\label{eq2}
&&\cH(\b^a, \pi_{a},P,Q) \nonumber \\
&= &\tilde{N} \left[\frac12 \ G^{ab} \pi_a \pi_b  +
\sum_A c_A (Q,P,\partial\b,\partial^2 \b, \partial Q)\exp\big(- 2 w_A (\b)\big)\right] \, .
\end{eqnarray}
Here $\pi_{a}$ (with $a = 1, ... ,d$) denote the canonical momenta conjugate to the ``logarithmic scale factors'' $\beta^a$, while $Q$ denote the remaining configuration variables ($\theta^a_i$, 3-form components $A_{ijk} (t, \textbf{x})$ in supergravity), and $P$ their canonically conjugate momenta ($P^i_a, \pi^{ijk}$). The symbol $\partial$ denotes {\em spatial}  derivatives.  The (inverse) metric $G^{ab}$ in Eq.~(\ref{eq2}) is the DeWitt ``superspace'' metric induced on the $\beta$'$s$ by the Einstein-Hilbert action. It endows the  $d$-dimensional\footnote{10 dimensional for SUGRA$_{11}$; but the various superstring theories also lead to a 10 dimensional Lorentz space   because one must add the (positive) kinetic term of the dilaton $\varphi \equiv \beta^{10} $ to the 9-dimensional DeWitt metric corresponding to the 9 spatial dimensions.} $\beta$ space with a Lorentzian structure
$ G_{ab} \, \dot{\beta}^a \dot{\beta}^b$. 

One of the crucial features of Eq.~(\ref{eq2}) is the appearance of Toda-like exponential potential terms 
$\propto \exp (- 2 w_A(\beta))$, where the $w_A(\beta)$ are {\em linear forms} in the logarithmic scale factors: $w_A(\beta) \equiv w_{Aa} \, \beta^a$. The range of labels $A$ and the specific ``wall forms'' $w_A(\beta)$ that appear depend on the considered model. For instance, in SUGRA$_{11}$ there appear: ``symmetry wall forms'' $w_{ab}^S (\beta) \equiv \beta^b - \beta^a$ (with $a < b$), ``gravitational wall forms'' $w_{abc}^g (\beta) \equiv 2\beta^a + \underset{e \neq a,b, c}{\sum} \,  \beta^e$ ($a \neq b$, $b \neq c$, $c \neq a$), ``electric 3-form wall forms'', $e_{abc} (\beta )
\equiv \beta^{a} + \beta^{b} + \beta^{c}$ ($a \neq b$, $b\neq c$, $c\neq a$), and ``magnetic 3-form wall forms'', $m_{a_1 .... a_6} \equiv \beta^{a_1} + \beta^{a_2} + ... + \beta^{a_6}$ (with indices all different).

One then finds that the near-spacelike-singularity limit amounts to considering the {\em large $\beta$ limit} in Eq.(2). In this limit a crucial role is played by the linear forms $w_A ( \beta)$ appearing in the ``exponential walls''. Actually, these walls enter in successive ``layers''. A first layer consists of a subset of all the walls called the {\em dominant walls} $w_i (\beta)$. The effect of these dynamically dominant walls is to confine the motion in $\beta$-space to a {\em fundamental billiard chamber} defined by the inequalities $w_i (\beta) \geqslant 0$. In the case of SUGRA$_{11}$, one finds that there are 10 dominant walls: 9 of them are the symmetry walls  $w_{12}^{S}(\beta), w_{23}^{S}(\beta), ... , w_{910}^{S}(\beta), $ and the 10th is an electric 3-form wall  $e_{123} (\beta) = \beta^1 + \beta^2 + \beta^3$. As noticed in \cite {hep-th/0012172} a remarkable fact is that the fundamental cosmological billiard chamber of SUGRA$_{11}$ (as well as type-II string theories) is the {\em Weyl chamber} of the hyperbolic Kac-Moody algebra $\E$. More precisely, the 10 dynamically dominant wall forms $\big\{w_{12}^{S}(\beta), w_{23}^{S}(\beta), ... , w_{910}^{S}(\beta), e_{123} (\beta)\big\}$  can be identified with the 10 {\em simple roots}    $ \{ \alpha_1(h), \alpha_2(h), ... , \alpha_{10}(h)\}$ of $E_{10}$. Here $h$ parametrizes a generic element of a Cartan subalgebra (CSA) of $E_{10}$ . [Let us also note that for Heterotic and type-I string theories the cosmological billiard is the Weyl chamber of another rank-10 hyperbolic Kac-Moody algebra, namely $B E_{10}$]. In the Dynkin diagram of $E_{10}$, Fig. 1, the 9 ``horizontal'' nodes correspond to the 9 symmetry walls, while the characteristic ``exceptional'' node sticking out ``vertically'' corresponds to the electric 3-form wall $e_{123} = \beta^1 + \beta^2 + \beta^3$.  [The fact that this  node stems from the {\em 3rd} horizontal node is then seen to be directly related to the presence of the {\it 3-form} $A_{\mu\nu\lambda}$, with electric kinetic energy $ \propto g^{i \ell} g^{jm} g^{kn} \dot{A}_{ijk}\dot{A}_{\ell m n}$].

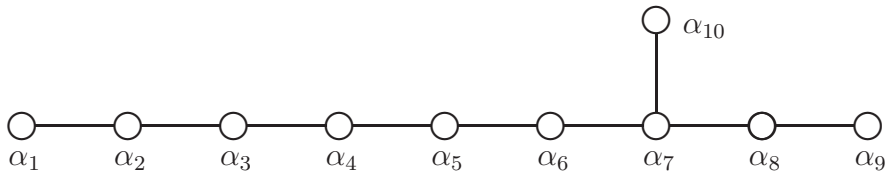
\begin{figure}[h]
\begin{center}
\scalebox{1}{
\begin{picture}(340,60)
\put(5,-5){$\alpha_1$}
\put(45,-5){$\alpha_2$}
\put(85,-5){$\alpha_3$}
\put(125,-5){$\alpha_4$}
\put(165,-5){$\alpha_5$}
\put(205,-5){$\alpha_6$}
\put(245,-5){$\alpha_7$}
\put(285,-5){$\alpha_8$}
\put(325,-5){$\alpha_9$}
\put(260,45){$\alpha_{10}$}
\thicklines
\multiput(10,10)(40,0){9}{\circle{10}}
\multiput(15,10)(40,0){8}{\line(1,0){30}}
\put(250,50){\circle{10}} \put(250,15){\line(0,1){30}}
\put(290,10){\circle{10}}
\end{picture}}
\caption{\label{e10dynk}\sl Dynkin diagram of $E_{10}$.}
\end{center}
\end{figure}

The  appearance of $\E$ in the BKL behaviour of SUGRA$_{11}$ revived an old suggestion of B.~Julia \cite{Julia85} about the possible role of $\E$ in a {\em one-dimensional reduction } of SUGRA$_{11}$. A posteriori, one can view the BKL behaviour as a kind of spontaneous reduction to one dimension (time) of a multidimensional theory. Note, however, that we are always discussing generic {\em inhomogeneous} 11-dimensional solutions, but that we examine them in the near-spacelike-singularity limit where the spatial derivatives are sub-dominant: $\partial_x \ll \partial_t$. Note also that the discrete $E_{10} (\mathbb{ Z})$ was proposed as a $U$-duality group of the  full $(T^{10})$ spatial toroidal compactification of $M$-theory by Hull and Townsend \cite{hep-th/9410167}.

\section{Gravity/Coset correspondence}

Refs \cite{hep-th/0207267,hep-th/0410245} went beyond the leading-order BKL analysis just recalled by including the first three ``layers'' of spatial-gradient-related sub-dominant walls $\propto \exp (- 2 w_A (\beta))$ in Eq.(\ref{eq2}).  The relative importance of these sub-dominant walls, which modify the leading billiard dynamics defined by the 10 dominant walls $w_i(\beta)$, can be ordered   by means of an expansion which counts how many dominant wall forms $w_i(\beta)$ are contained in the exponents of the sub-dominant wall forms $w_A(\beta)$, associated  to {\em higher spatial gradients}. By mapping the dominant gravity wall forms $w_i(\beta)$ onto the corresponding $\E$ simple roots $\a_i (h), i = 1, ..., 10,$ the just described BKL-type {\em gradient expansion} becomes mapped onto a Lie-algebraic {\it height expansion} in the roots of $\E$.  It was remarkably found that, up to height 30 (i.e. up to small corrections to the billiard dynamics associated to the product of 30 leading walls $e^{- 2 w_i (\beta) }$), the SUGRA$_{11}$ dynamics for $g_{\mu\nu} (t, \textbf{x})$, $A_{\mu\nu\lambda} (t, \textbf{x})$ considered  at some given spatial point $\textbf{x}_0$, could be identified to the geodesic dynamics of a {\em massless particle} moving on the (infinite-dimensional) coset space $\KE$. Note the ``holographic'' nature of this correspondence between an 11-dimensional dynamics on one side, and a 1-dimensional one on the other side.

A point on the coset space $E_{10}(\mathbb{R}) / K(E_{10}(\mathbb{R}))$ is coordinatized by a time-dependent (but spatially independent) element of the $E_{10}(\mathbb{R})$ group of the (Iwasawa) form: $g(t) = \exp h(t) \exp \nu (t)$. Here, $ h(t) = \beta_{\textrm{coset}}^a (t) H_a$ belongs to the 10-dimensional CSA of $E_{10}$, while $\nu (t) = \sum_{\alpha > 0} \nu^\alpha (t) E_\alpha$ belongs to a Borel subalgebra of $E_{10}$ and has an infinite number of components labelled by a {\em positive root} $\alpha $ of $E_{10}$.  The (null) geodesic action over the coset space $\KE$ takes the simple form
\begin{equation}
\label{eq3}
S_{E_{10}/K(E_{10})} = \int \frac{dt}{n(t)} (v^{\textrm{sym}} | v^{\textrm{sym}})
\end{equation}
where $v^{\textrm{sym}} \equiv \frac{1}{2} (v + v^T)$ is the ``symmetric''\footnote{Here the transpose operation $T$ denotes the negative of the Chevalley involution $\omega$ defining the real form  $E_{10(10)}$ of $E_{10}$. It is such that the elements $k$ of the Lie sub-algebra of $\K$ are ``$T$-antisymmetric'': $k^T = - k $, which is equivalent to them being fixed under $\omega : \omega(k)= + \, \omega(k)$.} part of the ``velocity''  $v \equiv (dg/dt) g^{-1}$ of a group element $g(t)$ running over $E_{10}(\mathbb{R})$.

The correspondence between the gravity, Eq.~(\ref{eq2}), and coset, Eq.~(\ref{eq3}), dynamics is best exhibited by decomposing (the Lie algebra of) $E_{10}$ with respect to (the Lie algebra of) the $GL(10)$ subgroup defined by the horizontal line in the Dynkin diagram of $E_{10}$. This allows one to grade the various components of $g(t)$ by their $GL(10)$ level $\ell$. One finds that, at the $\ell = 0$ level, $g(t)$ is parametrized by the Cartan coordinates $\beta_{\rm coset}^a (t)$ together with a unimodular upper triangular zehnbein $\theta_{{\rm coset} \, i}^a (t)$. At level $\ell = 1$, one finds a 3-form $A_{ijk}^{\rm coset} (t)$; at level $\ell = 2$, a 6-form $A_{i_1 i_2 \ldots i_6}^{\rm coset} (t)$, and at level $\ell = 3$ a $9$-index object $A_{i_1 \mid i_2 \ldots i_9}^{\rm coset} (t)$ with Young-tableau symmetry $\{ 8,1 \}$. The coset action (\ref{eq3}) then defines a coupled set of equations of motion for $\beta_{\rm coset}^a (t)$, $\theta_{{\rm coset} \, i}^a (t)$, $A_{ijk}^{\rm coset} (t)$, $A_{i_1 \ldots i_6}^{\rm coset} (t)$, $A_{i_1 \mid i_2 \ldots i_9}^{\rm coset} (t)$. By explicit calculations, it was found that these coupled equations of motion could be identified (modulo terms corresponding to potential walls of height at least 30) to the SUGRA$_{11}$ equations of motion, considered at some given spatial point $\textbf{x}_0$. 

The {\it dictionary} between the two dynamics says essentially that: 

\noindent (0) $\beta_{\rm gravity}^a (t,\textbf{x}_0) \leftrightarrow \beta_{\rm coset}^a (t) \, , \ \theta_i^a (t, \textbf{x}_0) \leftrightarrow \theta_{{\rm coset} \, i}^a (t)$, (1) $\partial_t \, A_{ijk}^{\rm coset} (t)$ corresponds to the electric components of the 11-dimensional field strength $F_{\rm gravity}$ $= d \, A_{\rm gravity}$ in a certain frame $e^i$, (2) the conjugate momentum of $A_{i_1 \ldots i_6}^{\rm coset} (t)$ corresponds to the {\it dual} (using $\varepsilon^{i_1 i_2 \ldots i_{10}}$) of the ``magnetic'' frame components of the 4-form $F_{\rm gravity} = d \, A_{\rm gravity}$, and (3) the conjugate momentum of $A_{i_1 \mid i_2 \ldots i_9} (t)$ corresponds to the $\varepsilon^{10}$ dual (on $jk$) of the structure constants $C_{jk}^i$ of the coframe $e^i$ ($d \, e^i = \frac{1}{2} \, C_{jk}^i \, e^j \wedge e^k$).

The fact that at levels $\ell = 2$ and $\ell = 3$ the dictionary between supergravity and coset variables maps the {\it first spatial gradients} of the SUGRA variables $A_{ijk} (t,\textbf{x})$ and $g_{ij} (t,\textbf{x})$ onto (time derivatives of) coset variables suggested the conjecture \cite{hep-th/0207267} of a hidden {\it equivalence} between the two models, i.e. the existence of a dynamics-preserving map between the infinite tower of (spatially independent) coset variables $(\beta_{\rm coset}^a , \nu^{\alpha})$, together with their conjugate momenta $(\pi_a^{\rm coset} , p_{\alpha})$, and the infinite sequence of spatial Taylor coefficients $(\beta (\textbf{x}_0)$, $\pi (\textbf{x}_0)$, $Q (\textbf{x}_0)$, $P (\textbf{x}_0)$, $\partial Q (\textbf{x}_0)$, $\partial^2 \beta (\textbf{x}_0)$, $\partial^2 Q(\textbf{x}_0), \ldots , \partial^n Q (\textbf{x}_0) , \ldots)$ formally describing the dynamics of the gravity variables $(\beta (\textbf{x}) , \pi (\textbf{x}) , Q(\textbf{x})$, $P(\textbf{x}))$ around some given spatial point $\textbf{x}_0$.\footnote{One, however, expects the map between the two models to become spatially non-local for heights $\geq 30$.}

It has been possible to extend the correspondence between the two models to the inclusion of fermionic terms on both sides \cite{hep-th/0512163,hep-th/0512292,hep-th/0606105}. Moreover, Ref. \cite{hep-th/0504153} found evidence for a nice compatibility between some high-level contributions (height $-115$!) in the coset action, corresponding to {\it imaginary} roots\footnote{i.e. such that $(\alpha , \alpha) < 0$, by contrast to the ``real'' roots, $(\alpha , \alpha) = +2$, which enter the checks mentionned above.}, and $M$-theory {\it one-loop corrections} to SUGRA$_{11}$, notably the terms quartic in the curvature tensor. (See also \cite{Damour:2006ez} for a study of the compatibility of an underlying Kac-Moody symmetry with quantum corrections in various models).

\section{A new view of the (quantum) fate of space at a cosmological singularity}

Let us now, following \cite{DNGRF07}, sketch the physical picture suggested by the gravity/coset correspondence. That is, let us take seriously the idea that, upon approaching a spacelike singularity, the description in terms of a spatial continuum, and space-time based (quantum) field theory breaks down, and should be replaced by a purely abstract Lie algebraic description. More precisely, we suggest that the information previously encoded in the spatial variation of the geometry and of the matter fields gets transferred to an infinite tower of spatially independent (but  time dependent) Lie algebraic variables. In other words, we are led to the conclusion that space actually ``disappears'' (or ``de-emerges'') as the singularity is approached\footnote{We have in mind here a ``big crunch'', i.e. we conventionally consider that we are tending {\it towards} the singularity. {\it Mutatis mutandis}, we would say that space ``appears'' or ``emerges'' at a big bang.}. In particular (and this would be bad news for Gabriele's pre-big bang scenario), we suggest no (quantum) ``bounce'' from an incoming collapsing universe to some outgoing expanding universe. Rather it is suggested that ``life continues'' for an infinite ``affine time'' at a singularity, with the double understanding, however, that: (i) life continues only in a totally new form (as in a kind of ``transmigration''), and (ii) an infinite affine time interval (measured, say, in the coordinate $t$ of Eq.~(\ref{eq3}) with a coset lapse function $n(t) = 1$) corresponds to a sub-Planckian interval of geometrical proper time\footnote{Indeed, it is found that the coset time $t$ (with $n(t) = 1$) corresponds to a ``Zeno-like'' gravity coordinate time (with rescaled lapse $\tilde N = N / \sqrt g = 1$) which tends to $+\infty$ as the proper time tends to zero.}.

Let us also comment on some expected aspects of the ``duality'' between the two models. It seems probable (from the AdS/CFT paradigm) that, even if the equivalence between the ``gravity'' and the ``coset'' descriptions is formally exact, each model has a natural domain of applicability in which the corresponding description is sufficiently ``weakly coupled'' to be trustable as is, even in the leading approximation. For the gravity description this domain is clearly that of curvatures smaller than the Planck scale. One then expects that the natural domain of validity of the dual coset model would correspond (in gravity variables) to that of curvatures larger than the Planck scale. In addition, it is possible that the coset description should primarily be considered as a quantum model, as now sketched.

The coset action (\ref{eq3}) describes the classical motion of a massless particle on the symmetric space $E_{10} ({\mathbb R}) / K (E_{10} ({\mathbb R}))$. Quantum mechanically, one should consider a quantum massless particle, i.e., if we neglect polarization effects\footnote{Actually, Refs. \cite{hep-th/0512163,hep-th/0512292,hep-th/0606105} indicate the need to consider a {\it spinning} massless particle, i.e. some kind of Dirac equation on $E_{10} / K (E_{10})$.} a Klein-Gordon equation,
\begin{equation}
\label{eq4}
\square \, \Psi (\beta^a , \nu^{\alpha}) = 0 \, ,
\end{equation}
where $\square$ denotes the (formal) Laplace-Beltrami operator on the infinite-dimensional Lorentz-signature curved coset manifold $E_{10} ({\mathbb R}) / K (E_{10} ({\mathbb R}))$. \break Eq.~(\ref{eq4}) would apply to the case considered here of un-compactified $M$-theory. In the case where all spatial dimensions are toroidally compactified, it has been suggested \cite{hep-th/9903110,hep-th/0401053} that $\Psi$ satisfy (\ref{eq4}) together with a condition of periodicity over the discrete group $E_{10} ({\mathbb Z})$. In other words, $\Psi$ would be a ``modular wave form'' on $E_{10} ({\mathbb Z}) \backslash E_{10} ({\mathbb R}) / K (E_{10} ({\mathbb R}))$.

Let us emphasize (still following \cite{DNGRF07}) that all reference to space and time has disappeared in Eq.~(\ref{eq4}). The disappearance of time is common between (\ref{eq4}) and the usual Wheeler-DeWitt equation in which the ``wave function(al) of the universe'' $\Psi [g_{ij} (\textbf{x})]$ no longer depends on any {\it extrinsic} time parameter. [As usual, one needs to choose, among all the dynamical variables a specific ``clock field'' to be used as an {\it intrinsic} time variable parametrizing the dynamics of the remaining variables.] The interesting new feature of (\ref{eq4}) (when compared to a Wheeler-DeWitt type equation) is the disappearance of any notion of geometry $g_{ij} (\textbf{x})$ and its replacement by the infinite tower of Lie-algebraic variables $(\beta^a , \nu^{\alpha})$\footnote{Note that this is conceptually very different from the $E_{11}$-based proposal of \cite{hep-th/0104081}.}. This quantum de-emergence of space, and the emergence of an infinite-dimensional symmetry group $E_{10}$\footnote{Let us note that $E_{10}$ enjoys a similarly distinguished status among the (infinite-dimensional) {\it hyperbolic} Kac-Moody Lie groups as $E_8$ does in the Cartan-Killing classification of the {\it finite-dimensional} simple Lie groups \cite{Kac}.} which {\it deeply intertwines space-time with matter degrees of freedom} might be radical enough to get us closer to an understanding of the fate of space-time and matter at cosmological singularities.

\vspace{1cm}

\noindent {\bf Acknowledgments.} It is a pleasure to dedicate this review to Gabriele Veneziano, a dear friend and a great physicist from whom I have learned a lot. I am also very grateful to my collaborators Marc Henneaux and Hermann Nicolai for the (continuing) $E_{10}$ adventure. I wish also to thank Maurizio Gasperini and Jnan Maherana for their patience.

\end{document}